\pdfoutput=1
\documentclass[twocolumn,showpacs,floatfix,prl]{revtex4}

\usepackage[final]{graphicx}
\DeclareGraphicsExtensions{.pdf}
\usepackage{amsmath,amssymb,bbold,bm}
\usepackage{float}

\newcommand{\bk}{{\bf k}}
\newcommand{\bp}{{\bf p}}

\newcommand{\bq}{{\bf q}}
\newcommand{\bmm}{{\bf m}}
\newcommand{\bn}{{\bf n}}

\newcommand{\bQ}{{\bf Q}}
\newcommand{\br}{{\bf r}}

\newcommand{\Tr}{{\rm Tr}}
\newcommand{\cF}{{\cal F}}
\newcommand{\cG}{{\cal G}}

\def\id{\mathbb{1}}

\begin{document}

\title{Theory of quasiparticle interference on the surface of a strong topological insulator}

\author{H.-M. Guo$^{1,2}$ and M. Franz$^{1}$}
\affiliation{$^{1}$Department of Physics and Astronomy, University
of British Columbia,Vancouver, BC, Canada V6T 1Z1}
\affiliation{$^{2}$Department of Physics, Capital Normal University,
Beijing 100048, China}

\begin{abstract}
Electrons on the surface of a strong topological insulator, such as
Bi$_2$Te$_3$ or Bi$_{1-x}$Sb$_x$, form a topologically protected
helical liquid whose excitation spectrum contains an odd number of
massless Dirac fermions. A theoretical survey and classification is
given of the universal features, observable by the ordinary and
spin-polarized scanning tunneling spectroscopy, in the interference
patterns resulting from the quasiparticle scattering by magnetic and
non-magnetic impurities in such a helical liquid. Our results
confirm the absence of backscattering from non-magnetic impurities
observed in recent experiments and predict new interference
features, uniquely characteristic of the helical liquid, when the
scatterers are magnetic.
\end{abstract}

\maketitle

\begin{figure}
\includegraphics[width=8cm]{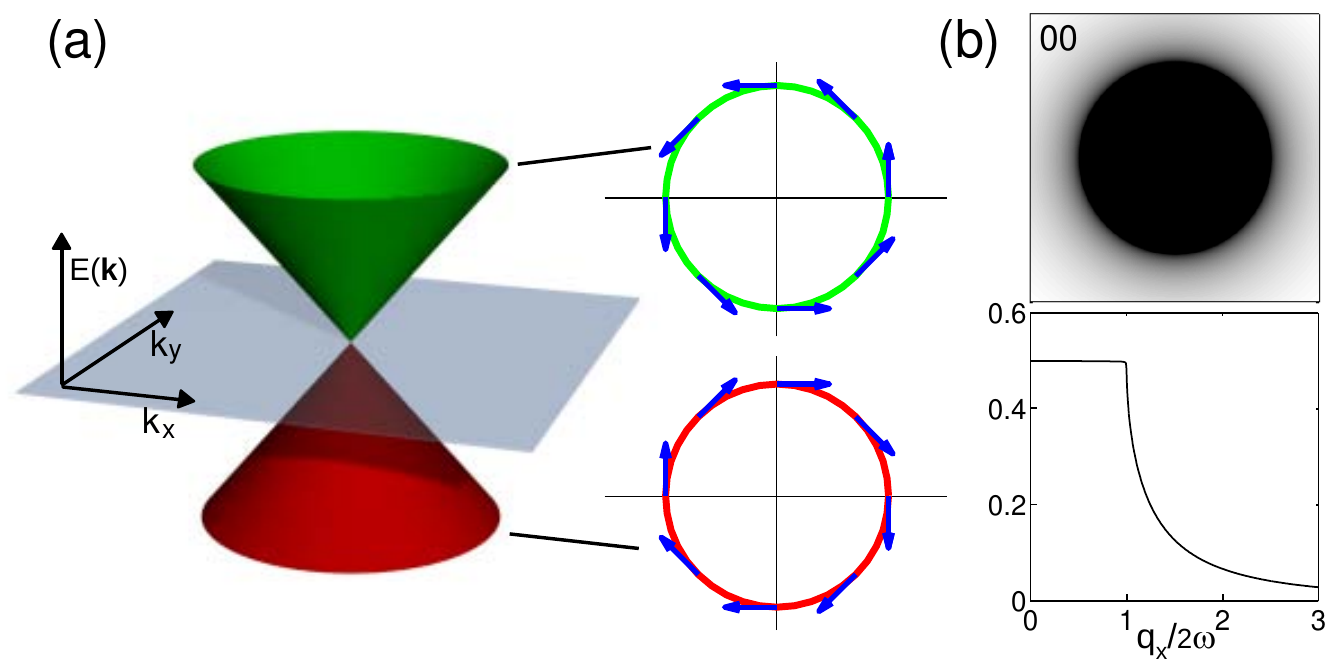}
\caption{(Color online) a) Dirac dispersion and spin orientation in
helical liquid on the surface of a strong topological insulator. b)
FT-STS response $\Im \Lambda_{00}(\bq,\omega)$ as a density plot in
the $(q_x,q_y)$ plane (top), cut along the $q_x$ direction
(bottom).} \label{fig1}
\end{figure}

A surface of the three dimensional strong topological insulator (STI)
\cite{mele1,moore1,qi1} is a very special place. Topological
invariants that characterize the bulk band structure of the underlying
time-reversal invariant crystal guarantee the existence of an odd
number of gapless surface states with the characteristic Dirac
dispersion. The surface electron spins are aligned in the plane of the
surface and point in the direction perpendicular to the momentum
vector, as indicated in Fig.\ \ref{fig1}a. Such arrangement of electron
spins and momenta has been termed `helical liquid' and is predicted to
exhibit a number of unusual physical properties
\cite{fu1,ran1,qi0,seradjeh1,qi2,raghu1}.

One important property of electrons in the helical liquid, noted
early on \cite{mele1}, is the absence of backscattering in the
presence of non-magnetic impurities. Since electrons with opposite
momenta also have opposite spins, backscattering requires a
spin-flip process, which cannot occur in the absence of
time-reversal symmetry ($\cal T$) breaking. This fundamental
property of the helical liquid has been recently tested in a series
of experiments \cite{yazdani1,manoharan1,kapitulnik1,xue1} using the
technique of the Fourier-transform scanning tunneling spectroscopy
(FT-STS) applied to the previously discovered STIs Bi$_{1-x}$Sb$_x$
and Bi$_2$Te$_3$ \cite{cava1,cava2}. The experimental results are
consistent with simple heuristic arguments for electron scattering
in the helical liquid.

In this Communication we develop a detailed theory of FT-STS in a
helical liquid formed on the surface of a STI. We focus on the
characteristic universal physics encoded in the low-energy massless
Dirac Hamiltonian \cite{mele1,raghu1},
\begin{equation}\label{h1} H=v{\bm
\sigma}\cdot(\hat{z}\times\bp)+\sum_{\alpha=0}^3\sigma_\alpha
V_\alpha(\br)
\end{equation}
describing the surface state of {\em all} STIs at probe frequencies $\omega$
tuned sufficiently close to the Dirac point. In Eq.\ (\ref{h1}) $v$ is
the Fermi velocity, ${\bm \sigma}=(\sigma_1,\sigma_2,\sigma_3)$ is the
vector of Pauli spin matrices, $\sigma_0=\id$,
$\bp=-i(\partial_x,\partial_y)$ is the planar momentum operator (we
take $\hbar=1$) and $V_\alpha(\br)$ is the impurity potential in
channel $\alpha$.
\begin{table*}
\caption{Results for FT-STS response function $2\pi\Lambda_{\alpha\beta}(\bq,i\omega)$ evaluated from
Eq.\ (\ref{lam2}). Here $\Omega$ represents the ultraviolet cutoff,
comparable to the bandwidth of the surface state, $z=2i\omega/|\bq|$
and $\hat{\bq}=(q_x,q_y)/|\bq|$. Functions $\cF(z)$ and $\cG(z)$ are
defined in the text. }
\begin{tabular}{|c||c|c c c|} \hline
$\alpha \setminus \beta$ & 0
& 1 & 2 & 3 \\ \hline\hline 0 & $\ln(1+\Omega^2/\omega^2)-\cF(z)$ &
$0$ & $0$ & $0$ \\
\hline
1 & $0$ & ~\ $\hat{q}_x^2[2+z^2\cG(z)]-1$ ~\
& $\hat{q}_x\hat{q}_y[2+z^2\cG(z)]$ & $-\hat{q}_xz\cG(z)$ \\
2 &
$0$ & $\hat{q}_x\hat{q}_y[2+z^2\cG(z)]$ & ~\
$\hat{q}_y^2[2+z^2\cG(z)]-1$ ~\ & $-\hat{q}_yz\cG(z)$ \\
3 & $0$
& $\hat{q}_xz\cG(z)$ & $\hat{q}_yz\cG(z)$ &
$-\ln(1+\Omega^2/\omega^2)-\cG(z)$ \\ \hline
\end{tabular}
\end{table*}
We study the ordinary tunneling and, with an eye on the future
experiments, we make predictions for the spin-resolved FT-STS
\cite{meier1}. We find that non-magnetic impurities produce only
weak, non-singular response in the ordinary FT-STS (Fig.\
\ref{fig1}b), consistent with the expected absence of
backscattering. Interestingly, we find that weak magnetic impurities
produce {\em no} response in the ordinary FT-STS, while strong
magnetic impurities produce only weak, non-singular response, except
possibly at a resonant frequency. The behavior is even more
interesting for a spin-resolved probe. Here, consistent with time
reversal symmetry, non-magnetic impurities produce no signal,
irrespective of their strength. Magnetic impurities, on the other
hand, give rise to distinctive FT-STS patterns with inverse square
root singularities at momenta $|\bq_\omega|=2\omega/v$. Depending on
the direction of the impurity magnetic moment and the direction in
which the probe spin current is polarized we find and classify
patterns with pronounced two- and four-fold rotational modulations.
Such modulated patterns, in conjunction with future spin-resolved
FT-STS experiments, can be used to probe the fundamental properties
of the helical liquid as well as the nature of disorder that
underlies these patterns. We note that other aspects of magnetic
impurities on STI surface have been studied in Ref.\ \cite{qi2}.

The power of FT-STS technique, originally developed in the context of
simple metals \cite{crommie1} and perfected in the studies of
high-temperature cuprate superconductors \cite{davis1}, lies in its
ability to use impurities, present in any real material, to probe the
electronic response of the underlying ideal {\em clean} material at finite momenta $\bq$. The
experiment measures the local density of states, $n(\br,\omega)$, at a
large number ($\sim 10^6$) of real-space locations $\br$ on the sample
surface. The spatial Fourier transform of this signal $n(\bq,\omega)$
contains the useful information. Theoretically it can be related
to the full electron propagator $G(\br,\br';\omega)$ as
\begin{equation}\label{n1}
n(\bq,\omega)=-{1\over \pi}\Im\int d^2r e^{-i\br\cdot\bq}\Tr[G(\br,\br;\omega)].
\end{equation}
Here the trace is taken over spin and  $\Im$ denotes the
strength of a branch cut across the real frequency axis $\Im
f(\omega)\equiv [f(\omega+i\delta)-f(\omega-i\delta)]/2i$, with
$\delta$ a positive infinitesimal.
In momentum space the electron propagator has a simple representation in terms of the $\hat{T}$ matrix,
\begin{equation}\label{g1}
G(\bk,\bk';\omega)=\delta_{\bk\bk'}G^0(\bk,\omega)+G^0(\bk,\omega)\hat{T}_{\bk\bk'}(\omega)G^0(\bk',\omega),
\end{equation}
where $G^0(\bk,\omega)=[\omega-v(k_x\sigma_y-k_y\sigma_x)]^{-1}$ is
the unperturbed propagator and the $\hat{T}$ matrix is subject to the Lippman-Schwinger equation
\begin{equation}\label{t1}
\hat{T}_{\bk\bk'}(\omega) = \hat{V}_{\bk-\bk'} + \sum_\bq
\hat{V}_{\bk-\bq} G^0(\bq,\omega)\hat{T}_{\bq\bk'}(\omega).
\end{equation}
Matrix $\hat{V}_\bk=\sum_\beta V^\beta_\bk\sigma_\beta$ is the Fourier
transform of the impurity term in Hamiltonian (\ref{h1}).

Below, in addition to the ordinary FT-STS we shall consider also the
spin-resolved FT-STS which is obtained by replacing  $\Tr[\sigma_i
G(\br,\br;\omega)]$ in the integrand of Eq.\ (\ref{n1}). Here and
hereafter Greek indices run from 0 to 3 while Roman indices run from
1 to 3. It is useful to consider the quantity
$n_{\alpha\beta}(\bq,\omega)$ which describes the FT-STS response of
a tunneling probe in charge/spin channel $\alpha$ to weak impurities
in charge/spin channel $\beta$.

When the
impurity potential is weak it is permissible to employ the Born
approximation, $\hat{T}_{\bk\bk'}(\omega) \approx
\hat{V}_{\bk-\bk'}$. In this limit, as first noted by Capriotti {\em
et al.} \cite{capriotti1}, the interesting $\bq$-dependent part of the
FT-STS signal can be expressed in a simple factorized form,
\begin{equation}\label{born1}
\delta n_{\alpha\beta}(\bq,\omega) =-{1\over \pi} |V^{\beta}_\bq|
{\Im} [\Lambda_{\alpha\beta}(\bq,\omega)],
\end{equation}
where
\begin{equation}\label{lam1}
\Lambda_{\alpha\beta}(\bq,\omega) =\sum_\bk\Tr[\sigma_\alpha G^0(\bk,\omega)\sigma_\beta G^0(\bk-\bq,\omega)].
\end{equation}
Since $V^{\beta}_\bq$ is a
Fourier transform of a random potential one expects it to be a
featureless function of $\bq$. $\Lambda_{\alpha\beta}(\bq,\omega)$, on
the other hand, represents the response of the underlying {\em clean}
system and contains, in general, prominent features as a function of
$\bq$ that can be used to directly infer its properties. We now
evaluate this response function for our model Hamiltonian
(\ref{h1}). Working in the continuum limit and passing to the
Matsubara frequency we have
\begin{equation}\label{lam2} \Lambda_{\alpha\beta}(\bq,i\omega)
=\int{d^2k\over (2\pi^2)} {L_{\alpha\beta}\over
[\omega^2+k^2][\omega^2+(\bk-\bq)^2]},
\end{equation}
where we also set $v=1$.  The factor in the numerator contains the
trace of the Pauli matrices and can be expressed as
$L_{\alpha\beta}=2\eta_\beta
\kappa_i\kappa'_j\Tr[\sigma_\alpha\sigma_i\sigma_\beta\sigma_j]$,
where ${\bm \kappa}=(k_x,k_y,i\omega)$, ${\bm
\kappa}'=(k_x-q_x,k_y-q_y,-i\omega)$ and $\eta_{0,3}=1$,
$\eta_{1,2}=-1$.

Integrals of the type indicated in Eq.\ (\ref{lam2}) are evaluated
most conveniently using the technique of Feynman parametrization
\cite{peskin} and have been studied in the context high-$T_c$ cuprate
superconductors \cite{tami1} and more recently graphene
\cite{tami3}. In fact we may read off the results for $\Lambda_{00}$
and $\Lambda_{33}$ directly from these studies and the remaining
integrals can be evaluated in a similar fashion. Our
results are summarized in Table I which represents the main result of
the present study. They are expressed in terms of two functions,
\begin{equation}\label{cg1}
\cG(z)={2\over\sqrt{-z^2-1}}\arctan{1\over\sqrt{-z^2-1}},
\end{equation}
and $\cF(z)=(-z^2-1)\cG(z)$ of the dimensionless variable
$z=2i\omega/|\bq|$. The results listed in the Table I satisfy
\begin{equation}\label{lamsym}
\Lambda_{\alpha\beta}(\bq,\omega)=\Lambda_{\beta\alpha}(-\bq,\omega),
\end{equation}
a property that can be easily established from  Eq.\ (\ref{lam1}).

The first line in the table $(\alpha=0)$ describes the ordinary
spin-unpolarized FT-STS. The response to non-magnetic impurities
$\Lambda_{00}$, shown in Fig.\ \ref{fig1}b, is non-vanishing but
weak. In the ordinary metal such response would entail a singularity
at momenta $\bq_\omega$ such that the band energy
$\epsilon(\bq_\omega/2)=\omega$, the probe bias. This singularity
arises from impurity scattering between states at momenta
$\bq_\omega /2$ and $-\bq_\omega /2$. In the present case of helical
liquid such backscattering processes are prohibited by time-reversal
symmetry \cite{mele1,yazdani1,manoharan1,kapitulnik1,xue1} and
consequently $\Im\Lambda_{00}$ shows only a kink.  Interestingly, we
find that FT-STS response to {\em magnetic impurities},
$\Lambda_{0i}$, vanishes identically in the Born limit. In this case
$\cal T$ does not prohibit backscattering but the response
nevertheless vanishes due to the symmetry property of the integral
in Eq.\ (\ref{lam2}). Specifically, Eq.\ (\ref{lamsym}) implies that
$\Lambda_{0i}(\bq,\omega)=\Lambda_{i0}(-\bq,\omega)$  and we shall
see below the that  latter must vanish due to ${\cal T}$-invariance.

Table I also indicates that response of spin-polarized FT-STS to
non-magnetic impurities, $\Lambda_{i0}$, vanishes. This can be
understood as follows. With non-magnetic impurities the surface of STI
remains $\cal T$-invariant and  therefore  cannot produce a response
in the spin channel. We expect this conclusion to remain valid even beyond the
Born approximation and beyond the simple linear Dirac model adopted in
this study.

\begin{figure}
\includegraphics[width=8.5cm]{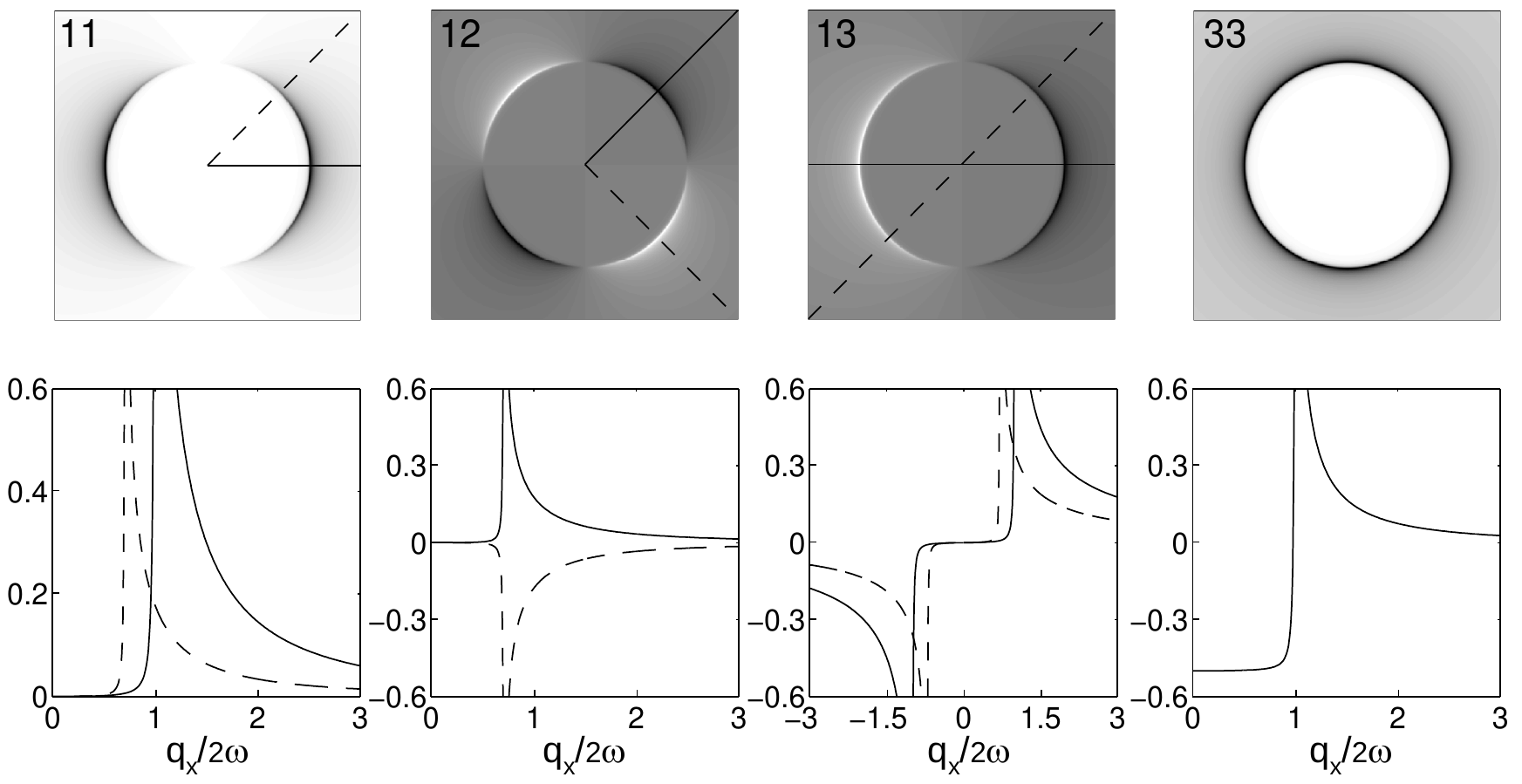}
\caption{Spin-resolved FT-STS response $\Im \Lambda_{ij}(\bq,\omega)$
  to magnetic impurities. Top row gives the density plot in the
  $(q_x,q_y)$ plane in channel $(ij)$ indicated in the upper right
  corner. Bottom row shows the corresponding cuts along the
  indicated lines parametrized by $q_x$.} \label{fig2}
\end{figure}

Spin-polarized FT-STS in the presence of magnetic impurities,
$\Lambda_{ij}$, shows the most interesting behavior. In all cases
$\Im\Lambda_{ij}(\bq,\omega)$ exhibits an inverse square-root
singularity, contained in the function $\cG(z)$ Eq.\ (\ref{cg1}), at
momenta $|\bq_\omega|=2\omega/v$. In addition, all channels except for
$\Lambda_{33}$, show interesting angular dependence on $\bq$ with
two- and  four-fold symmetry. For a general direction of impurity
magnetic moment $\hat{\bmm}$ and the STM tip spin-polarization
direction $\hat{\bn}$, the singular part of the response can be
written compactly as
\begin{equation}\label{lam7}
\Lambda^{\rm sing}_{\hat{\bmm}\hat{\bn}}(\bq,i\omega) = {1\over 2\pi}(\hat{\bmm}\cdot\bQ^*)(\hat{\bn}\cdot\bQ)z^2\cG(z),
\end{equation}
where $\bQ=(\hat{q}_x,\hat{q}_y,|\bq|/2i\omega)$ and $^*$ denotes complex conjugation.
These patterns, displayed in Fig.\
\ref{fig2}, could be used to identify the dominant source of
scattering in future spin-resolved FT-STS experiments. Specifically,
the distinctive angular dependence can be exploited to deduce the
direction of the magnetic moment of impurities deliberately deposited
on the surface of STI \cite{coral1}.

Our result of vanishing responses in $\Lambda_{0j}$ and $\Lambda_{i0}$
channels raises the question of whether this conclusion survives
beyond the Born approximation. Corrections to the Born approximations may
become important when the scattering potential is strong. To this end
we now consider the full $\hat{T}$ matrix governed by Eq.\ (\ref{t1}). For
an arbitrary impurity potential it is possible to find the solution
only numerically. For isolated point-like impurities it is however
possible to obtain exact analytical results. In the following we
shall focus on this limit.

For a single impurity characterized by a $\delta$-function real space
potential at the origin we have $\hat{V}_\bk=\sum_\beta V^\beta\sigma_\beta$,
independent of $\bk$. The solution for the $\hat{T}$ matrix is
simplest and most instructive when we consider scattering in a single
channel,  $\hat{V}_\bk=V^\beta\sigma_\beta$ (no summation
over $\beta$). The $\hat{T}$ matrix is then also momentum independent
and can be written as
\begin{equation}\label{t2}
\hat{T}_{\bk\bk'}^\beta(\omega) = \sum_\alpha\sigma_\alpha T_\alpha^\beta(\omega).
\end{equation}
Substituting this form into Eq.\ (\ref{t1}) and taking the trace we find
an equation for $\hat{T}_\alpha^\beta(\omega)$ of the form
\begin{equation}\label{t3}
T_\alpha^\beta(\omega) = V^\beta\delta_{\alpha\beta}+{1\over 2}V^\beta g_0(\omega)
\sum_\mu\Tr[\sigma_\alpha\sigma_\beta\sigma_\mu] T_\mu^\beta(\omega),
\end{equation}
where
\begin{equation}\label{gg0}
g_0(\omega)={1\over 2}\sum_\bk \Tr[G^0(\bk,\omega)]=-S{\omega\over
  4\pi}\ln\left(1-{\Omega^2\over\omega^2}\right).
\end{equation}
Here $S$ is the area of the STI surface and for finite concentration of
impurities it should be replaced by the inverse impurity density $n_I^{-1}$.
The solution  of Eq.\ (\ref{t3}) reads
\begin{eqnarray}\label{t4}
T_\beta^\beta(\omega) &=&{V^\beta\over 1-V^\beta g_0(\omega)}, \\
T_0^j(\omega) &=&{(V^j)^2g_0(\omega)\over 1-[V^j g_0(\omega)]^2}, \label{t5}
\end{eqnarray}
and all other components vanish.

Eq.\ (\ref{t4}) describes the
expected resonant enhancement of the scattering potential due to the
higher order terms in the Born series. This will affect the
amplitude but not the momentum structure of the FT-STS signal.

The result in Eq.\ (\ref{t5}) is more interesting and informs us that
starting at the second order in the Born expansion scattering potential in
the $\sigma_j$ channel (i.e.\ magnetic impurities with the moment along
direction $j$) produces non-zero $\hat{T}$ matrix in the charge
channel. Magnetic impurities, therefore, will be visible by the ordinary
FT-STS probe, although  the signal might generally be weak due to its appearance
in the second order of the expansion in the powers of $V^j$. Since the
$\hat{T}$ matrix
does not introduce any new momentum dependence for point-like scatterers
the momentum-space structure of $\delta n_{0j}(\bq,\omega)$ will be the
same as $\delta n_{00}(\bq,\omega)$, displayed in Fig.\ \ref{fig1}b.

Finally it is worth noting that Eq.\ (\ref{t3}) implies
$T_i^0(\omega)=0$, confirming our previously stated expectation that
non-magnetic impurities cannot produce a signal in the spin-resolved
FT-STS. Therefore, $\delta n_{i0}(\bq,\omega)$ remains zero to all
orders in the Born expansion, as dictated by ${\cal T}$-invariance of
the underlying system.

Our results presented above underscore the unique capabilities of the
FT-STS technique and its spin-polarized counterpart, applied to the surface of a
3-dimensional strong topological insulator. The expected absence of
backscattering in the helical liquid formed on such a surface in the
presence of non-magnetic
impurities \cite{mele1,yazdani1,manoharan1,kapitulnik1,xue1} is reflected by the
non-singular FT-STS response shown in Fig.\ \ref{fig1}a. This stands
in a sharp contrasts to the response of the ordinary metal which
involves inverse square root singularity when $\epsilon(\bq/2)=\omega$.

In addition to this finding our study reveals several striking and
unexpected features when the impurities are magnetic. First, although
backscattering is not prohibited in this case, we find that the FT-STS
response nevertheless {\em vanishes identically} for weak scattering potential treated in the
Born approximation. The reason for this unexpected null result is the
symmetry property Eq.\ (\ref{lamsym}) of the response function
$\Lambda_{\alpha\beta}(\bq,\omega)$ which equates the relevant
response $\Lambda_{0j}(\bq,\omega)$ to
$\Lambda_{j0}(-\bq,\omega)$. The latter in turn underlies the
spin-polarized response to weak nonmagnetic disorder and must
therefore vanish in a ${\cal T}$-invariant system. We note that this
conclusion does not rely in any way on our assumption of the Dirac Hamiltonian
(\ref{h1}) but follows from the general requirement of time reversal
invariance and symmetry (\ref{lamsym}), and will remain valid for
arbitrary momenta away from the Dirac point. Indeed this result is
applicable to an arbitrary ${\cal T}$-invariant system and indicates
that  the response of ordinary FT-STS to magnetic part of
the the scattering potential will generally be weak as it appears only
in the second order of the Born expansion.

In order to get strong FT-STS response to magnetic disorder one must employ the
spin-resolved tunneling probe \cite{meier1}. In this case the inverse square root divergence
at $\epsilon(\bq/2)=\omega$ is recovered. In our Dirac model of STI surface
(\ref{h1}) this singular response is accompanied by a distinctive
pattern of rotational anisotropy summarized by Eq.\ (\ref{lam7}) which
appears when either $\hat{\bmm}$ or $\hat{\bn}$ have a component in
the plane of the sample surface.

To model the FT-STS response away from the Dirac point one must include
corrections to the Dirac Hamiltonian reflecting the underlying
band structure at larger momenta. Very recently, such terms have been
considered to model
the surface state band structure of Bi$_2$Te$_3$ and results in good agreement
with experiments have been obtained \cite{hu1,arovas1}. For other
topological insulators such effects will be different as they depend
on the details of the underlying crystal structure. Our results, on
the other hand, will remain applicable to any STI surface when the
probe bias is tuned sufficiently close to the Dirac point and
constitute the universal
low-energy theory of FT-STS on the topologically protected STI surface states.

\emph{Acknowledgment}.---  The authors benefitted from the discussions with I.~Garate and T.~Pereg-Barnea. Support for this work came from NSERC, CIfAR and The
China Scholarship Council.

\end{document}